\documentclass[12pt]{article}
\usepackage{amsmath}
\usepackage{latexsym}
\usepackage{bbm}
\usepackage{color}

\def\a{\alpha}

\def\g{\gamma}

\def\l{\lambda}
\def\m{\mu}

\def\L{\Lambda}

\def\IZ{\mathbbm Z}

\def\and{{\rm and}}

\def\ie{{\it i.e.,} }

\begin{document}
\vspace*{-.6in} \thispagestyle{empty}
\begin{flushright}
CALT-68-2627
\end{flushright}
\baselineskip = 18pt

\vspace{1.5in} {\Large
\begin{center}
STRING THEORY: PROGRESS AND PROBLEMS\end{center}} \vspace{.5in}

\begin{center}
John H. Schwarz
\\
\emph{California Institute of Technology\\ Pasadena, CA  91125, USA}
\end{center}
\vspace{1in}

\begin{center}
\textbf{Abstract}
\end{center}
\begin{quotation}
\noindent String theory builds on the great legacy of Yukawa and
Tomonaga: New degrees of freedom and control of the UV are two
important themes. This talk will give an overview of some of the
progress and some of the unsolved problems that characterize string
theory today. It is divided into two parts:
\begin{itemize}
\item Connecting String Theory to the Real World

\item Gauge Theory/String Theory Duality
\end{itemize}
Two other major subjects, which I will omit, are Black Holes in
String Theory and The Impact of String Theory on Mathematics.
\end{quotation}

\bigskip

\centerline{Yukawa--Tomonaga Centennial Symposium}

\newpage

\pagenumbering{arabic}

\section{Connecting String Theory to the Real World}

String theory became a hot subject in the mid-1980s when it became
clear that it might give a deeper understanding of the origins of
the standard model and a consistent quantum theory containing
gravity \cite{Green:1987sp}\cite{Polchinski:1998rq}\cite{Becker:2007}. 
At that time five consistent string theories
were known, each of which requires ten spacetime dimensions and
supersymmetry. They are called:

\centerline{\it Type I, Type IIA, Type IIB}

\centerline{\it $SO(32)$ heterotic, $E_8 \times E_8$ heterotic}

\noindent Each of these theories is entirely free of adjustable
dimensionless parameters. All dimensionless parameters arise in
either of two possible ways:
\begin{itemize}
\item dynamically as the expectation values of scalar fields

\item as integers that count something such as topological
invariants, physical objects (branes), or quantized fluxes.
\end{itemize}

\subsection{Calabi--Yau Compactification}

One scheme looked particularly promising in the mid-1980s.
Specifically, the $E_8 \times E_8$ heterotic theory has consistent
vacuum solutions in which six spatial dimensions form a compact
Calabi--Yau manifold, which has SU(3) holonomy, and the other four
dimensions form Minkowski spacetime \cite{Candelas:1985en}. Thus,
the ten-dimensional spacetime ${\cal M}_{10}$ is a direct product
$$
{\cal M}_{10} = {\rm CY}_6 \times M_{3,1}.
$$

The effective four-dimensional theories that characterize such
solutions at low energies have the following attractive features:

\begin{itemize}
\item They have the structure of supersymmetric
grand unified theories. The well-known advantages of low-energy
supersymmetry and grand unification are therefore naturally
incorporated.

\item Each solution has a definite number of families
of quarks and leptons determined by the topology of the CY space.

\item The standard model gauge symmetry is embedded in one $E_8$
factor, and there is a hidden sector, associated to the second $E_8$
factor. Supersymmetry can break dynamically (by gluino condensation)
in the hidden sector. This breaking is communicated gravitationally
to the visible sector. Such schemes suppress unwanted
flavor-changing neutral currents, though possibly not as much as is
required.

\item There are several good dark-matter candidates:
The lightest supersymmetric particle, called the LSP (perhaps a
neutralino) is absolutely stable if there is an unbroken R symmetry,
as generally supposed to be the case. A stable neutralino has the
right properties for weakly interacting cold dark matter (a WIMP).
Other possibilities include gravitinos, axions, and hidden-sector
particles.
\end{itemize}

Despite the exuberance that was in the air in 1985,\footnote{I felt
at the time that the subject had undergone an almost instantaneous
transition from being vastly under-appreciated to being
over-optimistically regarded as being on the verge of providing a
{\it theory of everything}.} there was much that was not yet
understood. The successes were qualitative, and there were many
problems and puzzling questions. Some of these problems and
questions were the following:

\begin{itemize}
\item Hundreds of Calabi--Yau manifolds were known (now there are many
thousands). Which one of them, if any, is the {\it right one}? Is
there a principle (other than agreement with observations) by which
the right one can be determined? Are there string-theory based
schemes other than CY compactification of the $E_8 \times E_8$
heterotic theory that can give quasi-realistic solutions?

\item Why are there four other consistent superstring theories?
After all, we only need one fundamental theory.  Are some of them
inconsistent, or else, could some of them somehow be equivalent?

\item The CY compactification scenario was analyzed
using perturbation theory, but there is no good reason to believe
that the string coupling should be small. What new nonperturbative
features appear at strong coupling? Does the same qualitative
picture continue to hold at strong coupling?

\item The CY solutions typically give many massless scalars (called
{\it moduli}). Since the moduli have gravitational strength
interactions, they are ruled out by standard tests of GR. How can we
get rid of them? The effective potential does not depend on the
values of the moduli, so they describe {\it flat directions}. One
should somehow stabilize the moduli by generating an effective
potential with isolated minima and no flat directions.

\item What ensures that the vacuum energy density
(or dark energy) is sufficiently small, namely of order $10^{-120}$
in Planck units? In 1985 it was generally believed to be zero, so
one popular idea was to look for a symmetry principle that would
enforce this. It is just as well that such a symmetry was never
found, since we now know that the vacuum energy density is not
exactly zero. A generic nonsupersymmetric vacuum is expected to give
a vacuum energy of order one. Supersymmetry, broken at the TeV
scale, cuts this down to $10^{-60}$, which is only half way to the
right value on a logarithmic scale.

\item What does string theory have to say about cosmology?
\end{itemize}

\subsection{Lessons of the Second Superstring Revolution}

In the mid-1990s there was a remarkable burst of progress in
addressing some of the issues listed above. The main lessons were
the following:

\begin{itemize}
\item There is just one theory!
What had been viewed as five theories are actually five different
(highly symmetric) corners of a space of solutions to a unique
underlying theory. The five superstring theories are related by
various dualities:

{\it T-duality ($R \to \ell_{\rm s}^2/R$)} relates the two type II
superstring theories, and also the two heterotic string theories.
Here, $R$ represents the radius of a circular extra dimension and
$\ell_{\rm s}$ is the fundamental string length scale. There are
analogous dualities involving pairs of CY manifolds. The two CY
spaces are said to be related by {\it mirror symmetry}.

\medskip

{\it S-duality ($g_{\rm s} \to 1/g_{\rm s}$)} relates the type I
superstring theory and the $SO(32)$ heterotic string theory. Here,
$g_{\rm s}$ denotes the dimensionless string coupling constant,
whose value is determined by the vacuum expectation value of a
certain scalar field called the {\it dilaton}. The type IIB
superstring theory has an $SL(2,\IZ)$ duality group, which contains
this transformation as a special case.

\medskip

The type I theory can be derived from the type IIB theory by a
procedure called {\it orientifold projection}. This projection
essentially enforces left-right symmetry on the world-sheet theory.

\item New objects, called $p$-branes, arise nonperturbatively.
Here $p$ is an integer that represents the number of spatial
dimensions of the brane. ({\it Brane} is a made-up word derived from
membrane.) Stable $p$-branes carry conserved charges and satisfy
generalized Dirac quantization conditions. There are also unstable
$p$-branes that do not carry conserved charges. The main categories
of stable $p$-branes are D-branes, M-branes, and NS5-branes.
NS5-branes are magnetic duals of fundamental strings in the
heterotic and type II theories. (Type I strings do not carry a
conserved charge, which is one way of understanding why they can
break.)

\item D-branes are characterized by Dirichlet boundary conditions for
open strings. Also, the stable D-branes in type I and type II
superstring theories carry conserved charges of the Ramond--Ramond
type \cite{Polchinski:1995mt}. The relation of D-branes to open
strings has the crucial consequence that the world-volume theories
of D$p$-branes are Yang--Mills gauge theories in $p+1$ dimensions.
$p$ takes even values in the type IIA theory and odd values in the
type IIB theory. A system of $N$ coincident D$p$-branes gives a
$U(N)$ gauge theory. Other gauge groups can also be obtained, if the
D-branes coincide with certain types of singularities.

\medskip

\item In units where the fundamental superstring has tension (energy
density) of the form $$T_{F1} \sim m_{\rm s}^2,$$ the D$p$-branes
have tension
$$
T_{Dp} \sim m_{\rm s}^{p+1}/g_{\rm s}.
$$
This is in contrast with more conventional nonperturbative
excitations, such as monopoles in gauge theory or NS5-branes in
superstring theory, which have mass or tension proportional to $
1/g^2$.

\item The strong-coupling limit of the Type IIA superstring theory or the
$E_8 \times E_8$ heterotic string theory gives 11-dimensional
M-theory. The size of the 11th dimension is
$$
R_{11} \sim g_{\rm s} \ell_{\rm s}.
$$
In the $E_8 \times E_8$ heterotic case the 11th dimension is a line
interval, so the 11-dimensional spacetime has two 10-dimensional
boundaries. One set of $E_8$ gauge fields is localized on each
boundary. This gauge group is singled out by quantum consistency
(absence of anomalies).

After Calabi--Yau compactification of the strongly coupled $E_8
\times E_8$ heterotic string theory, it is possible for the volume
of the CY manifold to vary along the 11th dimension, which opens up
new possibilities for phenomenology. This approach is referred to as
{\it heterotic M-theory}.

\end{itemize}

\subsection{Intersecting D-brane Models}

\parindent = 1.0cm

The fact that D-branes carry Yang--Mills gauge theories suggests
another approach to model-building -- this time based on type II
superstrings. One again starts with a compactification of the form
$K_6 \times M_{3,1}$. Then one introduces D-branes that fill the
four-dimensional spacetime. The way to do this is to introduce
D$(3+n)$-branes that wrap $n$-dimensional cycles of $K_6$, while the
other dimensions are spacetime filling.

For example, consider D6-branes in the type IIA theory compactified
on a six-torus. Suppose that a set of $N_1$ D6-branes wraps three
compact dimensions and a second set of $N_2$ D6-branes wraps a
different three-cycle. The two stacks of branes give a $U(N_1)\times
U(N_2)$ gauge theory in four dimensional spacetime. Open strings
connecting the two stacks of branes at their intersection points
give chiral matter transforming as $(N_1, N_2)$. This is roughly the
type of structure that one has in the standard model and many of its
possible extensions. If certain restrictions are satisfied, these
constructions can give supersymmetric theories.

There are many schemes of this type. The successes and difficulties
in this approach to model building are comparable to those of CY
compactification of the heterotic string.

\subsection{Flux compactifications}

Type II superstrings contain various massless antisymmetric tensor
gauge fields (in the RR sector). These are conveniently described as
differential forms
$$
A_n = \frac{1}{n!} A_{\m_1 \m_2 \ldots \m_n} dx^{\m_1} \wedge
dx^{\m_2} \wedge \ldots \wedge dx^{\m_n}.
$$
The integer $n$ is odd in the type IIA theory and even in the type
IIB theory. These have gauge-invariant field strengths of the form
$$F_{n+1} = dA_n,$$
generalizing Maxwell theory ($n=1$). D-branes are sources for these
fields. There is also a two-form $B_2$ (in the NS sector) for which
the fundamental strings are electric sources and NS5-branes are
magnetic sources.

If the compact dimensions contain nontrivial $n$-cycles, $C_n$, it
is possible to have (quantized) flux threading the cycle:
$$
\int_{C_n} A_n = 2\pi N.
$$
Such possibilities greatly increase the number of possible quantum
vacua, though there are constraints that must be satisfied. Mike
Douglas has analyzed one particular CY compactification of the type
IIB theory, and he has estimated the number of distinct flux vacua
to be about $10^{500}$ \cite{Douglas:2003um}.

\subsection{Warped Compactification}

One of the important properties of flux compactifications is that
they give rise to warped geometries. This means that the
ten-dimensional geometry is no longer a direct product of the form
$${\cal M}_{10} = K_6 \times M_{3,1}.$$
Instead, the 4d Minkowski metric part of the 10d metric is
multiplied by a warp factor $h(y)$ that depends on position $y$ in
the internal manifold
$$
ds^2_{10} = h(y) dx \cdot dx + ds^2_6.
$$
Also the internal six-manifold is no longer Calabi--Yau, since it is
also multiplied by a warp factor.

In the brane-world picture one can have spacetime-filling D3-branes
that are localized at points in the internal manifold. Randall and
Sundrum \cite{Randall:1999ee} have proposed that a large ratio
between the warp factors at the position of a standard model brane
and a Planck brane could provide a solution of the hierarchy
problem, accounting for the large ratio between the Planck scale and
the weak scale. The RS scenario can be made rather precise in the
context of flux compactifications with branes.

\subsection{Moduli Stabilization and the Landscape}

The moduli problem --- the occurrence of massless scalar fields
$\phi_i$ with continuously adjustable vacuum expectation values ---
can be solved in the context of flux compactification. The fluxes
induce a nontrivial potential energy function for the moduli
$V(\phi_i)$. This {\it landscape} has isolated minima, which are
what Douglas counted. The moduli are massive at a minimum. The
details are a bit complicated, as some moduli are stabilized by
perturbative effects and others are stabilized by nonperturbative
effects.

There are many issues. For example, one of the moduli fields is the
{\it dilaton} $\Phi$, whose vacuum value determines the string
coupling constant:
$$
g_{\rm s} = \langle e^\Phi \rangle.
$$
If this is stabilized at a nonzero value, then the system is
inherently nonperturbative, and perturbation theory (an expansion
about the free theory value $g_{\rm s} = 0$) does not make sense.
Even though the qualitative picture may survive, this is a serious
technical problem.

The proliferation of vacua raises many questions: Is it completely
hopeless to find the right one? Is it meaningful or useful to assign
probabilities to the vacua and study them statistically? One
proposal is that such a probability distribution is determined by
the wave function of the Universe, which might be determined by a
suitable initial value boundary condition.

\subsection{The Cosmological Constant}

The values of the cosmological constant in the various (metastable)
vacua seem to be more or less randomly distributed over a range of
order unity (in Planck units). If this is the case, then about 1 in
$10^{120}$ would have roughly the right value. If there really are
$10^{500}$ or more vacua, this is a very large number even though it
is a very small fraction.

Weinberg has argued that if the cosmological constant were more than
an order of magnitude or two larger than it is, galaxies wouldn't
form, and then we wouldn't be here to discuss the question. This
type of reasoning may be correct, but it leaves me with an
unsatisfied feeling. I hope that we can find the right vacuum, or
the right cosmological solution, and use it to answer many of the
questions that motivated us to become physicists.

Even if string theory provides the correct theory, and it is
completely unique, it is still a big question how predictive the
theory is. If the number of consistent vacua (or solutions) is too
large, and there is no principle for making an informed choice among
them, it might not be possible (even in principle) to compute
fundamental constants such as the fine-structure constant and the
electron--muon mass ratio. However, I remain optimistic that this is
not the case.

\subsection{Brane Inflation}

There is a lot of evidence that supports the hypothesis that the
very early Universe underwent a period of inflation during which the
scale factor grew exponentially by a factor of at least $e^{60}$
(\ie 60 e-foldings). In simple field theory models this is described
in terms of a ``slowly rolling'' scalar field called an {\it
inflaton}.

The subject of string cosmology has become very active in the past
several years. Proposals for the string-based origins of inflation,
or possible alternatives, are being explored extensively. Let me
sketch a specific scenario due to Kachru et al. \cite{Kachru:2003aw}
\cite{Kachru:2003sx}. It takes place in the setup of CY
compactification with flux and warped throats in the geometry.

Inflation takes place as a D3-brane moves down a throat, attracted
to an anti-D3-brane at the bottom until they collide and annihilate.
A scalar mode of an open string connecting the branes is the
inflaton. The annihilation releases the tension energy stored in the
branes. It heats up the Universe to start the hot big-bang epoch.
All sorts of strings are produced, and some might survive to be
observable as {\it cosmic superstrings}. Their tensions could be
small enough to have avoided detection so far, due to the warping of
the geometry. So, while there can be no definitive prediction at
this time, the search for linear lenses in the sky is one way in
which fundamental strings might be discovered.

\section{Gauge Theory/String Theory Duality}

A class of dualities --- referred to as {\it AdS/CFT dualities} or
{\it holographic dualities} --- was proposed by Maldacena
\cite{Maldacena:1997re}. A very symmetrical example is the duality
between ${\cal N} = 4$ supersymmetric Yang--Mills theory in four
dimensions \cite{Brink:1976bc}, with gauge group $SU(N)$,  and type
IIB superstring theory \cite{Green:1981yb} in an $AdS_5 \times S^5$
background \cite{Schwarz:1983qr} with $N$ units of RR flux. $S^n$
denotes an $n$-dimensional sphere, and $AdS_n$ refers to anti de
Sitter space, a maximally symmetric spacetime geometry with negative
curvature, in $n$ dimensions. This geometry is certainly
unrealistic, but it is an excellent theoretical laboratory.

One class of generalizations relates various superconformal field
theories in four dimensions to type IIB superstring configurations
with spacetime backgrounds of the form $AdS_5 \times X_5$. Here
$X_5$ is a Sasaki--Einstein space, which has the property that a
six-dimensional cone over $X_5$ is a noncompact Calabi--Yau space.
Another class of generalizations introduces deformations such that
the gauge theory no longer has conformal symmetry, and the dual
$AdS_5$ geometry is deformed in a corresponding manner. Therefore,
the subject of gauge theory/string theory duality entails much more
than AdS/CFT duality.

\subsection{An Example}

The discussion that follows is restricted to the $AdS_5 \times S^5$
example. The most obvious test of the duality is to check that both
theories have the same symmetry. The symmetry in each case is known
to be the supergroup $PSU(2,2|4)$, so this test is passed.
Exhibiting the dynamical details of the duality is much more
challenging for a number of reasons. For one thing, the gauge theory
is a nontrivial interacting field theory. However, it is somewhat
special: there are indications that it may be integrable in the
planar approximation. The planar approximation corresponds to the
leading terms in the 't Hooft large-$N$ expansion
\cite{'tHooft:1973jz}, which is carried out for fixed 't Hooft
parameter
$$
\l = g_{\rm YM}^2 N.
$$
The large-$N$ expansion is dual to the string-theory perturbation
expansion, since the string coupling constant $g_{\rm s}$ is given
by
$$
g_{\rm s} = \frac{g_{\rm YM}^2}{4\pi}  = \frac{\l}{4\pi N}.
$$

Even in the planar approximation, in which the strings do not
interact, the string theory analysis is very challenging. The
spectrum of string excitations is easy to compute in Minkowski
spacetime, but very hard in a curved spacetime geometry. The string
world-sheet theory (a two-dimensional QFT) is believed to be
integrable in the case of $AdS_5 \times S^5$. The geometry (and
hence the string spectrum) depends on the dimensionless parameter
$\a'/R^2$, where $\a' = \ell_{\rm s}^2$ is the string Regge slope
parameter and $R$ is the radius of the $S^5$ and the $AdS_5$. This
ratio corresponds to $1/\sqrt{\l}$. Thus, the supergravity
approximation ($\a' \ll R^2$) corresponds to large 't Hooft coupling
$\l$.

There is a lot of effort underway trying to solve the gauge theory
in the planar approximation, as well as the string world-sheet
theory, and to understand how they are related. Specifically, one
wants to relate the anomalous dimensions of gauge-invariant
operators in the gauge theory to the energies of type IIB string
excitations. Much ingenious work has been done, but this seems to be
a very challenging program.

An important aspect of the characterization of these operators and
excitations is the amount of supersymmetry that they possess. The
ground state preserves all of the supersymmetry. Beyond that, there
are states (or operators) that preserve one-half, one-quarter,
one-eighth, one-sixteenth or none of the supersymmetry. As the
amount of supersymmetry is decreased, the complexity of the analysis
increases. The states that preserve half of the supersymmetry, for
example, are fully understood from the gauge theory perspective
\cite{Corley:2001zk}\cite{Berenstein:2004kk}. 
They are also understood from the type IIB
superstring theory perspective in a supergravity/semiclassical
approximation \cite{Lin:2004nb}.

\subsection{D-brane Model of QCD}

As mentioned earlier, an interesting approach to particle physics
phenomenology is based on intersecting D-brane models. There are a
lot of issues to contend with, and it is difficult to address them
all at once. A more modest approach than trying to find a system
that accounts for all known interactions in a realistic way is to
isolate a subset of the features that one would like to implement
and analyze how they can be achieved.

A good example is to restrict to hadron physics and try to construct
a D-brane model of QCD. This brings one back to the original goal of
string theory research in the late 1960s and early 1970s -- the
construction of a theory of the strong interactions. But now we can
approach the problem with many more tools than were available at
that time.

$N_c$ coincident D-branes give an $SU(N_c)$ gauge theory, so that is
a good start. One way to break supersymmetry is to take one
dimension to be a circle of radius $R$ and give fermions
antiperiodic boundary conditions on the circle. If one choose
D4-branes (in the type IIA theory) that wrap the circle, following
Witten \cite{Witten:1998zw}, then their five-dimensional
world-volume theory reduces to pure four-dimensional gauge theory at
energies $E\ll 1/R$. the dual geometry can be treated in a
supergravity approximation for large $N_c$ and large 't Hooft
coupling.

The addition of quark-like matter to this system was considered by
Sakai and Sugimoto \cite{Sakai:2004cn}. Their proposal is to add
$N_f$ coincident D8-branes and $N_f$ coincident anti-D8-branes that
are localized on the circle. These D-branes are introduced using a
probe approximation. This means that their effect on the spacetime
geometry created by the D4-branes is neglected. The D4-D8 open
strings give $N_f$ quark flavors in the gauge theory. At low
energies this results in a Born--Infeld type action in a curved
background.

The gauge symmetry of the D8-branes results in a
$$
U(N_f)_L \times U(N_f)_R
$$
global chiral symmetry in the 4d gauge theory. As is well-known from
studies of QCD, this should break spontaneously to a vectorlike
$SU(N_f)$ subgroup with the appearance of adjoint massless
Nambu--Goldstone bosons (pions).

Studying the BI equations of motion, SS show that the branes
reconnect into a single set of D8-branes in a U-like configuration.
This beautifully accounts for the required {\it chiral-symmetry
breaking} and gives the requisite {\it massless pions} as internal
components of the higher-dimensional gauge fields. Moreover, the
effective low-energy theory of the pions has the correct structure.

This construction captures the physics of QCD correctly at low
energies, where the pion Lagrangian is the whole story. However, it
differs from QCD at higher energies. The key unphysical feature is
that the QCD scale $\L$ is comparable to the Kaluza--Klein scale
$1/R$. In a more realistic scheme the dual geometry would have
string-scale curvature, and therefore it would not admit a
supergravity approximation. This is one reason why the string theory
dual of QCD is not yet known. At least we now know that it involves
more than four dimensions.

\subsection{Application of AdS/CFT to RHIC Physics}

RHIC collides gold nuclei with 200 GeV per nucleon. This presumably
produces a quark-gluon plasma (QGP) with $T\approx 250$ MeV and
$\a_s \approx 0.5$. This is a strongly coupled nonabelian plasma.
The collisions produce an enormous number of particles, which makes
it difficult to extract significant physics results. One way of
probing of what is happening is to observe the rate and spectrum of
D-meson production. From this one infers how the energy of a charm
quark is dissipated as it moves through the hot plasma. This is
characterized by the {\it viscous drag}.

In order to utilize AdS/CFT to study the strongly coupled theory, {
one replaces QCD by ${\cal N} = 4$ SYM}. This sounds like a terrible
thing to do, but it is probably not as bad for high temperature as
it would be at zero temperature. (At high temperature, both theories
both are deconfined, for example.) The temperature is taken account
in the dual picture by replacing $AdS_5$ with an {\it $AdS_5$ black
hole} with Hawking temperature $T$. Since $\l \sim 10$, a
supergravity approximation can be used.

Bulk thermodynamic quantities have a finite limit for $\l \to \infty$.
In particular, the dimensionless ratio
$$
\frac{\rm shear \ viscosity}{\rm entropy \ density} \to
\frac{1}{4\pi}
$$
was computed by Policastro, Son, Starinets \cite{Policastro:2001yc}.
This very low value is in rough agreement with deductions based on
measurements of the {\it elliptic flow} of collisions. The viscous
drag
$$
\frac{dp}{dt} = -\frac{\pi}{2} \sqrt{\l} T^2 \frac{v}{\sqrt{1-v^2}},
$$
obtained by analyzing open strings connecting a D7-brane to the
black hole, also gives rough agreement with observations.

There are many uncertainties in relating theory and experiment, but
I don't think there is a better approach on the market. I find it
quite amazing that some nuclear physicists have become interested in
black holes in five-dimensional anti de Sitter space!

\subsection{Matching Critical Exponents}

In 1993 Choptuik discovered by numerical studies that if one forms a
black hole by gravitational collapse of a free massless scalar field
and varies the conditions of the collapse by changing some control
parameter $f$, there is a critical value of $f$ above which a black
hole forms. The mass of the resulting black hole scales as
$$
M_{\rm BH} \sim (f - f_c)^\g.
$$
Choptuik found that $\g_4 \approx .372$ in four dimensions. Sorkin
and Oren \cite{Sorkin:2005vz} found that in five dimensions
$$\g_5 = 0.408 \pm 2\%.$$

The dual gauge theory process is the high energy Regge domain, which
is dominated by Pomeron exchange --- the BFKL
(Balitsky-Fadin-Kuraev-Lipatov) Pomeron, to be precise. One-Pomeron
exchange violates the unitarity bound above a certain energy (since
the Regge intercept is $\a(0) >1$), but this is cured at high energy
by the nonlinear evolution of the BFKL kernel, which takes account
of multi-Pomeron effects. This is characterized by a critical
exponent
$$
\g_{\rm BFKL} = 0.409552,
$$
This is in striking agreement with $\g_5 = 0.408 \pm 2\%$
\cite{Alvarez-Gaume:2006dw}. It will be interesting to obtain a more
accurate computation of $\g_5$ and see whether this agreement
continues to hold.\footnote{Note added: A recent computation 
gives $\g_5 = 0.4131 \pm .0001$, which seems to indicate 
that the two numbers are not the same \cite{Bland:2007sg}.}

\section{Conclusion}

There has been a lot of progress in understanding string theory and
its possible connections to the real world, but many problems
remain. Even if progress continues to be made at the current rapid
pace, I do not expect the subject to be completely understood by the
time of the Yukawa--Tomonaga bicentennial. I do not consider this to
be a pessimistic viewpoint. After all, it would be sad if we were so
successful that we put ourselves out of business.

I am grateful to the organizers for having had the opportunity to
participate in this centennial conference. This work was partially
supported by the U.S. Dept. of Energy under Grant No.
DE-FG03-92-ER40701.


\bigskip

\begin{thebibliography}{99}

%\cite{Green:1987sp}
\bibitem{Green:1987sp}
  M.~B.~Green, J.~H.~Schwarz and E.~Witten,
  {\it Superstring Theory}, Vols.  I and II, Cambridge Univ. Press
  (1987).

%\cite{Polchinski:1998rq}
\bibitem{Polchinski:1998rq}
  J.~Polchinski,
  {\it String Theory}, Vols. I and II, Cambridge Univ. Press
  (1998).

%\cite{Becker:2007}
\bibitem{Becker:2007}
  K. Becker, M. Becker and J.~H.~Schwarz,
  {\it String Theory and M-Theory: A Modern Introduction},  Cambridge Univ. Press
  (2007).

%\cite{Candelas:1985en}
\bibitem{Candelas:1985en}
  P.~Candelas, G.~T.~Horowitz, A.~Strominger and E.~Witten,
  %``Vacuum Configurations For Superstrings,''
  Nucl.\ Phys.\ B {\bf 258}, 46 (1985).
  %%CITATION = NUPHA,B258,46;%%

%\cite{Polchinski:1995mt}
\bibitem{Polchinski:1995mt}
  J.~Polchinski,
  %``Dirichlet-Branes and Ramond-Ramond Charges,''
  Phys.\ Rev.\ Lett.\  {\bf 75} (1995) 4724
  [arXiv:hep-th/9510017].
  %%CITATION = HEP-TH 9510017;%%

%\cite{Douglas:2003um}
\bibitem{Douglas:2003um}
  M.~R.~Douglas,
  %``The statistics of string / M theory vacua,''
  JHEP {\bf 0305}, 046 (2003)
  [arXiv:hep-th/0303194].
  %%CITATION = HEP-TH 0303194;%%

%\cite{Randall:1999ee}
\bibitem{Randall:1999ee}
  L.~Randall and R.~Sundrum,
  %``A large mass hierarchy from a small extra dimension,''
  Phys.\ Rev.\ Lett.\  {\bf 83}, 3370 (1999)
  [arXiv:hep-ph/9905221].
  %%CITATION = HEP-PH 9905221;%%

%\cite{Kachru:2003aw}
\bibitem{Kachru:2003aw}
  S.~Kachru, R.~Kallosh, A.~Linde and S.~P.~Trivedi,
  %``De Sitter vacua in string theory,''
  Phys.\ Rev.\ D {\bf 68}, 046005 (2003)
  [arXiv:hep-th/0301240].
  %%CITATION = HEP-TH 0301240;%%

%\cite{Kachru:2003sx}
\bibitem{Kachru:2003sx}
  S.~Kachru, R.~Kallosh, A.~Linde, J.~M.~Maldacena, L.~McAllister and S.~P.~Trivedi,
  %``Towards inflation in string theory,''
  JCAP {\bf 0310}, 013 (2003)
  [arXiv:hep-th/0308055].
  %%CITATION = HEP-TH 0308055;%%

%\cite{Maldacena:1997re}
\bibitem{Maldacena:1997re}
  J.~M.~Maldacena,
  %``The large N limit of superconformal field theories and supergravity,''
  Adv.\ Theor.\ Math.\ Phys.\  {\bf 2}, 231 (1998)
  [Int.\ J.\ Theor.\ Phys.\  {\bf 38}, 1113 (1999)]
  [arXiv:hep-th/9711200].
  %%CITATION = HEP-TH 9711200;%%

%\cite{Brink:1976bc}
\bibitem{Brink:1976bc}
  L.~Brink, J.~H.~Schwarz and J.~Scherk,
  %``Supersymmetric Yang-Mills Theories,''
  Nucl.\ Phys.\ B {\bf 121}, 77 (1977).
  %%CITATION = NUPHA,B121,77;%%

%\cite{Green:1981yb}
\bibitem{Green:1981yb}
  M.~B.~Green and J.~H.~Schwarz,
  %``Supersymmetrical String Theories,''
  Phys.\ Lett.\ B {\bf 109}, 444 (1982).
  %%CITATION = PHLTA,B109,444;%%

%\cite{Schwarz:1983qr}
\bibitem{Schwarz:1983qr}
  J.~H.~Schwarz,
  %``Covariant Field Equations Of Chiral N=2 D = 10 Supergravity,''
  Nucl.\ Phys.\ B {\bf 226}, 269 (1983).
  %%CITATION = NUPHA,B226,269;%%

%\cite{'tHooft:1973jz}
\bibitem{'tHooft:1973jz}
  G.~'t Hooft,
  %``A PLANAR DIAGRAM THEORY FOR STRONG INTERACTIONS,''
  Nucl.\ Phys.\ B {\bf 72}, 461 (1974).
  %%CITATION = NUPHA,B72,461;%%

%\cite{Corley:2001zk}
\bibitem{Corley:2001zk}
  S.~Corley, A.~Jevicki and S.~Ramgoolam,
  %``Exact correlators of giant gravitons from dual N = 4 SYM theory,''
  Adv.\ Theor.\ Math.\ Phys.\  {\bf 5}, 809 (2002)
  [arXiv:hep-th/0111222].
  %%CITATION = 00203,5,809;%%

%\cite{Berenstein:2004kk}
\bibitem{Berenstein:2004kk}
  D.~Berenstein,
  %``A toy model for the AdS/CFT correspondence,''
  JHEP {\bf 0407}, 018 (2004)
  [arXiv:hep-th/0403110].
  %%CITATION = HEP-TH 0403110;%%

%\cite{Lin:2004nb}
\bibitem{Lin:2004nb}
  H.~Lin, O.~Lunin and J.~M.~Maldacena,
  %``Bubbling AdS space and 1/2 BPS geometries,''
  JHEP {\bf 0410}, 025 (2004)
  [arXiv:hep-th/0409174].
  %%CITATION = HEP-TH 0409174;%%

%\cite{Witten:1998zw}
\bibitem{Witten:1998zw}
  E.~Witten,
  %``Anti-de Sitter space, thermal phase transition, and confinement in  gauge
  %theories,''
  Adv.\ Theor.\ Math.\ Phys.\  {\bf 2}, 505 (1998)
  [arXiv:hep-th/9803131].
  %%CITATION = HEP-TH 9803131;%%

%\cite{Sakai:2004cn}
\bibitem{Sakai:2004cn}
  T.~Sakai and S.~Sugimoto,
  %``Low energy hadron physics in holographic QCD,''
  Prog.\ Theor.\ Phys.\  {\bf 113}, 843 (2005)
  [arXiv:hep-th/0412141].
  %%CITATION = HEP-TH 0412141;%%

%\cite{Policastro:2001yc}
\bibitem{Policastro:2001yc}
  G.~Policastro, D.~T.~Son and A.~O.~Starinets,
  %``The shear viscosity of strongly coupled N = 4 supersymmetric Yang-Mills
  %plasma,''
  Phys.\ Rev.\ Lett.\  {\bf 87}, 081601 (2001)
  [arXiv:hep-th/0104066].
  %%CITATION = HEP-TH 0104066;%%

%\cite{Sorkin:2005vz}
\bibitem{Sorkin:2005vz}
  E.~Sorkin and Y.~Oren,
  %``On Choptuik's scaling in higher dimensions,''
  Phys.\ Rev.\ D {\bf 71}, 124005 (2005)
  [arXiv:hep-th/0502034].
  %%CITATION = HEP-TH 0502034;%%

%\cite{Alvarez-Gaume:2006dw}
\bibitem{Alvarez-Gaume:2006dw}
  L.~Alvarez-Gaume, C.~Gomez and M.~A.~Vazquez-Mozo,
  %``Scaling phenomena in gravity from QCD,''
  arXiv:hep-th/0611312.
  %%CITATION = HEP-TH 0611312;%%

%\cite{Bland:2007sg}
\bibitem{Bland:2007sg}
  J.~Bland and G.~Kunstatter,
  %``The 5-D Choptuik critical exponent and holography,''
  arXiv:hep-th/0702226.
  %%CITATION = HEP-TH/0702226;%%

\end{thebibliography}
\end{document}